\documentclass[11pt]{article}
\pdfoutput=1
\usepackage{fullpage}
\usepackage{xcolor}
\usepackage{cite}
\usepackage{graphicx}
\usepackage{tensor}
\usepackage{amsmath,bm}
\usepackage{amssymb}
\usepackage{subfig}
\usepackage[utf8x]{inputenc} 
\title{}
\date{}
\renewcommand{\vec}[1]{\mbox{\boldmath$ #1 $}}

\def\beq{\begin{equation}}
\def\eeq{\end{equation}}
\allowdisplaybreaks

\begin{document}
\bibliographystyle{utphys}

\newcommand{\hel}{\eta} 
\renewcommand{\d}{\mathrm{d}}
\newcommand{\dd}{\hat{\mathrm{d}}}
\newcommand{\del}{\hat{\delta}}
\newcommand{\ket}[1]{| #1 \rangle}
\newcommand{\bra}[1]{\langle #1 |}

\newcommand{\be}{\begin{equation}}
\newcommand{\ee}{\end{equation}}
\newcommand\n[1]{\textcolor{red}{(#1)}} 
\newcommand{\diff}{\mathop{}\!\mathrm{d}}
\newcommand{\lb}{\left}
\newcommand{\rb}{\right}
\newcommand{\f}{\frac}
\newcommand{\pd}{\partial}
\newcommand{\tr}{\text{tr}}
\newcommand{\fdiff}{\mathcal{D}}
\newcommand{\im}{\text{im}}
\let\caron\v
\renewcommand{\v}{\mathbf}
\newcommand{\T}{\tensor}
\newcommand{\R}{\mathbb{R}}
\newcommand{\C}{\mathbb{C}}
\newcommand{\Z}{\mathbb{Z}}
\newcommand{\msbar}{\ensuremath{\overline{\text{MS}}}}
\newcommand{\DIS}{\ensuremath{\text{DIS}}}
\newcommand{\abar}{\ensuremath{\bar{\alpha}_S}}
\newcommand{\bb}{\ensuremath{\bar{\beta}_0}}
\newcommand{\rc}{\ensuremath{r_{\text{cut}}}}
\newcommand{\Nd}{\ensuremath{N_{\text{d.o.f.}}}}
\newcommand{\red}[1]{{\color{red} #1}}

\newcommand{\Ad}{\dot{A}}
\newcommand{\Bd}{\dot{B}}
\newcommand{\Cd}{\dot{C}}
\newcommand{\Dd}{\dot{D}}
\newcommand{\Ed}{\dot{E}}
\newcommand{\Fd}{\dot{F}}
\newcommand{\depsilon}{\epsilon}
\newcommand{\dsigma}{\bar{\sigma}}

\newcommand{\bphi}{\phi} 
\newcommand{\bB}{B} 
\newcommand{\bH}{H} 
\newcommand{\bsigma}{\sigma} 
\newcommand{\charge}{\tilde{c}} 
\newcommand{\ampA}{\mathcal{A}} 
\newcommand{\ampM}{\mathcal{M}} 

\titlepage

\vspace*{0.5cm}

\begin{center}
{\bf \Large Classical strings and the double copy}

\vspace*{1cm} 
\textsc{Riccardo Morieri$^a$\footnote{riccardo.morieri@edu.unito.it}, Igor Pesando$^a$\footnote{igor.pesando@unito.it}, Michael L. Reichenberg Ashby$^b$\footnote{m.l.reichenbergashby@qmul.ac.uk}
  and Chris D. White$^b$\footnote{christopher.white@qmul.ac.uk}} \\

\vspace*{0.5cm} $^a$ Dipartimento di Fisica, Universit\`{a} di Torino, and INFN, Sezione di Torino, Via P. Giuria 1, I-10125 Torino;\\

\vspace*{0.5cm} $^b$ Centre for Theoretical Physics, School of Physical and
Chemical Sciences, \\ Queen Mary University of London, 327 Mile End
Road, London E1 4NS, UK\\

\end{center}

\vspace*{0.5cm}

\begin{abstract}
  The double copy is by now a well-established relationship between
  scattering amplitudes and classical solutions in gauge and gravity
  (field) theories, and is itself inspired by amplitude relations in
  string theory. In this paper, we generalise the classical double
  copy to the motion of strings, taking as a case study the motion of
  an open string in a background abelian gauge field. We argue that
  the double copy of this situation is a closed string moving in a
  spacetime background arising as the double copy of the gauge theory
  background. The gauge theory background we consider is that of a
  constant electric field, which has a critical value beyond which the
  open string motion is pathological. We find no counterpart of this
  behaviour in the double copy, and interpret this result. We then
  examine how the closed string nevertheless still knows about the
  single copy gauge theory. Our results pave the way for more
  systematic study of the double copy in a classical string context,
  thus going beyond the KLT relations for amplitudes in flat space.
\end{abstract}

\vspace*{0.5cm}

\section{Introduction}
\label{sec:intro}

The double copy is by now a well-established correspondence linking
quantities in different quantum field theories. Its original
incarnation was for non-abelian gauge and gravity theories, where it
related scattering amplitudes in flat
space~\cite{Bern:2008qj,Bern:2010ue,Bern:2010yg}. This was itself
inspired by the well-known KLT relations in string
theory~\cite{Kawai:1985xq}, which state that tree-level closed string
scattering amplitudes are equivalent to sums of products of their
open-string counterparts. At low energy, closed and open strings can
give rise to gravitons and gluons respectively, such that the KLT
relations and double copy are equivalent. The latter, however, has
been tested up to four-loop order depending on the field theory,
where a direct string explanation is lacking\footnote{For progress on
extending the KLT relations to loop level, see
e.g. refs.~\cite{Stieberger:2022lss,Stieberger:2023nol}.}. Thus,
whether or not the double copy has a true underlying string theory
explanation remains an open question.

Shortly after the double copy for scattering amplitudes appeared, it
was extended to certain exact classical solutions of gauge and gravity
theories~\cite{Monteiro:2014cda}. There are by now a wide range of
approaches for examining classical solutions, including those that can
only be treated order-by-order in perturbation
theory~\cite{Monteiro:2014cda,Luna:2015paa,Ridgway:2015fdl,Bahjat-Abbas:2017htu,Carrillo-Gonzalez:2017iyj,CarrilloGonzalez:2019gof,Bah:2019sda,Alkac:2021seh,Alkac:2022tvc,Luna:2018dpt,Sabharwal:2019ngs,Alawadhi:2020jrv,Godazgar:2020zbv,White:2020sfn,Chacon:2020fmr,Chacon:2021wbr,Chacon:2021hfe,Chacon:2021lox,Dempsey:2022sls,Easson:2022zoh,Chawla:2022ogv,Han:2022mze,Armstrong-Williams:2022apo,Han:2022ubu,Elor:2020nqe,Farnsworth:2021wvs,Anastasiou:2014qba,LopesCardoso:2018xes,Anastasiou:2018rdx,Luna:2020adi,Borsten:2020xbt,Borsten:2020zgj,Goldberger:2017frp,Goldberger:2017vcg,Goldberger:2017ogt,Goldberger:2019xef,Goldberger:2016iau,Prabhu:2020avf,Luna:2016hge,Luna:2017dtq,Cheung:2016prv,Cheung:2021zvb,Cheung:2022vnd,Cheung:2022mix,Chawla:2024mse,Keeler:2024bdt,Chawla:2023bsu,Easson:2020esh,Armstrong-Williams:2024bog,Armstrong-Williams:2023ssz,Farnsworth:2023mff}. Further
non-perturbative insights have been obtained in
refs.~\cite{Monteiro:2011pc,Borsten:2021hua,Alawadhi:2019urr,Banerjee:2019saj,Huang:2019cja,Berman:2018hwd,Alfonsi:2020lub,Alawadhi:2021uie,White:2016jzc,DeSmet:2017rve,Bahjat-Abbas:2018vgo,Armstrong-Williams:2025spu,Cheung:2022mix,Moynihan:2021rwh,Borsten:2022vtg}
(see
e.g. refs.~\cite{Borsten:2020bgv,Bern:2019prr,Adamo:2022dcm,Bern:2022wqg,White:2021gvv,White:2024pve}
for pedagogical reviews). Given this widening remit, and the fact that
the ultimate origin and scope of the double copy remain somewhat
mysterious, it is timely to consider other possible extensions of the
double copy idea, including connections to other relevant literature
from either field or string theories.

With the above motivation in mind, our aim in this paper is to explore
generalising the classical double copy from field to string
theory. Put another way, the traditional classical double copy relates
particle-like objects in gauge and gravity theory, where charge
degrees of freedom in the former are replaced by kinematic variables
(e.g. mass, momentum or energy) in the latter. If the double copy has
an underlying string theory motivation, as suggested by the KLT
relations for amplitudes in flat space, then it should indeed
generalise to situations involving classical string motion. One would
then seek to replace a given scenario involving open string motion,
with a suitable motion for closed strings. To this end, we will
consider the particular situation of an open string moving in a
background gauge field, with the double copy of this situation then
comprising a closed string moving in a background spacetime found by
double-copying the gauge background. Importantly, we will take the
background spacetime to have a Kerr--Schild form, such that the by now
well-established rules of the {\it Kerr--Schild double
  copy}~\cite{Monteiro:2014cda} allow us to relate the two
backgrounds. We will further restrict our attention to the case in
which the electric field in the gauge theory is constant. Although
somewhat simple, this has a number of features that are well-suited to
a first case study of a classical double copy for strings. Firstly,
both the gauge and gravity backgrounds can be shown to be exact
solutions of the full string equations of motion, thereby including
all-order dependence in the string tension $\alpha'$. Secondly, there
is a well-known instability for open strings moving in a constant
electric
field~refs.~\cite{Fradkin:1985qd,Burgess:1986dw,Nesterenko:1989pz},
beyond which the string tension is no longer sufficient to bind the
string. Given that the electric field specifically acts on the charged
endpoints of the string, there can then be no instability for the
closed string. We will see that this is confirmed by our double copy
analysis. However, we will see that the double copy of the background
gauge field picks out a non-inertial frame, whose acceleration is
directly related to the electromagnetic field in the gauge theory. A
coordinate transformation yields a Minkowski-like space, with
restricted coordinates due to a Rindler horizon. Hence, there is a
sense in which the double copy is always visible in the gravity
situation, either locally or through the boundary conditions. This in
turn makes contact with other classical double copy results, in which
pathologies in a gauge theory can map to non-trivial boundary
conditions in gravity~\cite{Moynihan:2025vcs}. Given the wealth of
results concerning classical string motion, we hope that our results
inspire similar case studies, which in turn may extend our knowledge
of the origin and scope of the double copy itself.

The structure of our paper is as follows. In section~\ref{sec:review},
we review various aspects of the classical double copy, and clarify a
technical issue that is needed for what follows. In
section~\ref{sec:gauge}, we introduce the gauge field corresponding to
a constant electric field, and obtain its double copy to gravity. In
section~\ref{sec:strings}, we interpret the motion of open and closed
strings in the gauge and gravity backgrounds respectively, paying
particular attention to how singular behaviour is or is not
manifest. Finally, we discuss our results and conclude in
section~\ref{sec:discuss}.

\subsection*{Note added}

During the final stages of this paper, ref.~\cite{Alencar:2026zdz} appeared, which also discusses classical string solutions in the framework of the Kerr--Schild double copy. Our results do not overlap with that paper, which discusses particular extended string solutions and how to interpret their single and zeroth copies. It will be interesting in future to see whether our results can be combined with those of ref.~\cite{Alencar:2026zdz} in further examining the interplay between string theory and the classical double copy. 

\section{The Kerr--Schild double copy}
\label{sec:review}

In this paper, we will focus on strings moving in classical background
fields, such that gravity background are obtained as classical double
copies of gauge theory counterparts. To this end, we will use the
canonical formulation of the classical double copy that was first
presented in refs.~\cite{Monteiro:2014cda}, and that relies on the
special class of so-called {\it Kerr--Schild solutions} in General
Relativity. The starting point is to consider the generic
decomposition of the full metric tensor $g_{\mu\nu}$ in terms of a
Minkowski background $\eta_{\mu\nu}$, and a graviton $h_{\mu\nu}$:
\begin{equation}
  g_{\mu\nu}=\eta_{\mu\nu}+\kappa h_{\mu\nu},\quad \kappa^2=\sqrt{32\pi G_N},
\label{gmunu}
\end{equation}
where $G_N$ is Newton's constant. Whilst this decomposition can always
be made, certain gravitons assume the Kerr--Schild form
\begin{equation}
  h_{\mu\nu}(x)=\phi(x)k_\mu(x)k_\nu(x),
  \label{hmunu}
\end{equation}
involving a scalar field $\phi(x)$ and null vector field $k_\mu(x)$,
where the latter satisfies the null and geodesic conditions
\begin{equation}
  \eta^{\mu\nu}k_\mu k_\nu=g^{\mu\nu}k_\mu k_\nu=0,\quad
  k^\mu\partial_\mu k^\nu=0.
\label{kKS}
\end{equation}
As is well-known (see e.g. ref.~\cite{Stephani:2003tm} for a textbook
treatment), the Kerr--Schild ansatz has the effect of linearising the
mixed-index Ricci tensor:
\begin{equation}
  {R^\mu}_\nu=\frac12 \left(
  \partial^\mu\partial_\alpha(\phi k^\alpha k_\nu)
  +\partial_\nu\partial^\alpha(\phi k_\alpha k^\mu)
  -\partial^2(\phi k^\mu k_\nu)
  \right).
  \label{Ricci}
\end{equation}
The Einstein equations (with mixed-index placement) in turn become
linearised, and thus easier to solve. Although abstract at first
glance, many of the most well-known solutions in General Relativity
can be cast in a Kerr--Schild form, including the Schwarzschild and
Kerr black holes. More generally, Kerr--Schild solutions typically have
Petrov type D~\cite{Stephani:2003tm}. Given a Kerr--Schild solution, we
can naturally construct a non-abelian gauge field according to 
\begin{equation}
  {\bf A}_\mu=A_\mu^a{\bf T}^a,\quad A_\mu^a=c^a\phi(x)k_\mu(x),
  \label{Amua}
\end{equation}
where ${\bf T}^a$ is a colour generator, and a constant colour vector
$c^a$ has replaced one of the kinematic vectors $k^\mu$ in the
original Kerr--Schild graviton. The resulting gauge field turns out to
linearise the Yang-Mills equations, such that we may ignore the
non-abelian dependence (i.e. the presence of $c^a$ and the adjoint
index $a$ on the gauge field itself) from now on. Furthermore,
ref.~\cite{Monteiro:2014cda} showed that for static solutions
\begin{equation}
  \partial_0\phi=\partial_0k_\mu=0
  \label{static}
\end{equation}
with $k_0=1$, the gauge field $A_\mu=\phi k_\mu$ solves the Maxwell
equations
\begin{equation}
  \partial_\mu F^{\mu\nu}=0,\quad F_{\mu\nu}=\partial_\mu A_\nu
  -\partial_\nu A_\mu.
  \label{Fmunu}
\end{equation}
Thus, $A_\mu$ is a well-defined gauge theory counterpart, or {\it
  single copy}, of the graviton $h_{\mu\nu}$. Further incarnations of
the classical double copy based on spinor and twistor
methods~\cite{Luna:2018dpt,White:2020sfn,Chacon:2021wbr,Luna:2022dxo}
have made clear that the classical double copy is precisely related to
the original BCJ double copy for scattering amplitudes of
refs.~\cite{Bern:2008qj,Bern:2010ue,Bern:2010yg}.

We will label Cartesian coordinates in natural units according to
\begin{equation}
  x^\mu=(t,x,y,z)
  \label{xmudef},
\end{equation}
and for what follows it is also convenient to introduce the lightcone
coordinates
\begin{equation}
  u=\frac{t+x}{\sqrt{2}},\quad v=\frac{t-x}{\sqrt{2}},
  \label{lightcone}
\end{equation}
such that the Minkowski line element in both coordinate systems is
\begin{equation}
  ds^2=-dt^2+dx^2+dx_i dx^i=-2dudv+dx_i dx^i.
  \label{ds2Mink}
\end{equation}
Here, we have separated the coordinate $x$ from the other spatial
coordinates $\{x^i\}$, $i\geq 2$, motivated by the fact that we
will be concerned with gauge fields associated with a particular
direction in space, which we can take to be aligned with the
$x$-axis. The Kerr--Schild vectors entering such gauge fields can then
be chosen to be of form
\begin{equation}
  k=a(x^\lambda) du,
  \label{kudef}
\end{equation}
where the geodesic condition further implies
\begin{displaymath}
  k^\mu\partial_\mu k_\nu= k^0\partial_0 a+k^x\partial_x a+k^i\partial_i a=
  k^x\partial_x a=0,
\end{displaymath}
and thus $a\equiv a(x^i)$. Including also the scalar $\phi(x^\mu)$,
the most general static gauge field with such a Kerr--Schild vector is
\begin{equation}
  A=A_u(x,x^i)du.
  \label{Audef}
\end{equation}
It is interesting to ask whether there is any residual gauge freedom
that preserves the form of the Kerr--Schild vector. To find this, one
may write the gauge-transformed field as
\begin{equation}
  \tilde{A}_\mu=\phi k_\mu+\partial_\mu\theta=
  \phi\left(k_\mu+\frac{\partial_\mu \theta}{\phi}\right)=
  \phi\tilde{k}_\mu,
  \label{Atildemu}
\end{equation}
and then demand that the the new Kerr--Schild vector satisfies
\begin{equation}
  \tilde{k}_\mu=\tilde{a}(x^i)du.
  \label{ktilde}
\end{equation}
This in turn implies
\begin{equation}
  \partial_v\theta=\partial_i\theta=0,
  \label{ktilde2}
\end{equation}
such that we may take $\theta\equiv\theta(u)$. Comparing
eqs.~(\ref{kudef}, \ref{ktilde}) we must also have
\begin{equation}
  \tilde{a}(x^i)=a(x^i)+\frac{\theta'(u)}{\phi},
  \label{atilde1}
\end{equation}
which implies
\begin{equation}
  \theta'(u)=G(x,x^i)
  \label{theta'}
\end{equation}
for some function $G(x,x^i)$. The only way this can be satisfied is if
both sides are equal to a constant, so that the possible gauge
parameters are restricted to
\begin{equation}
  \theta(u)=\alpha u.
  \label{thetasol}
\end{equation}
This amounts to a constant shift in the $u$ component of the gauge
field.

\section{Constant electric fields and their double copy}
\label{sec:gauge}

Having reviewed general aspects of the Kerr--Schild double copy in the
previous section, we now turn to a specific example of gauge fields
associated with the null direction of eq.~(\ref{kudef}), namely gauge
fields that gives rise to a constant electromagnetic field. We note
further that such cases are different to the plane waves already
considered in ref.~\cite{Monteiro:2014cda} (see also
refs.~\cite{Godazgar:2020zbv,Bahjat-Abbas:2020cyb}), as we are now
seeking solutions with an extended profile in the $x$-direction,
rather than having this confined to a plane at fixed $u$. Our aim is
to explain and interpret the double copy of constant electromagnetic
fields, as a precursor to analysing string motion in these backgrounds
in the following section.

Before considering the special case of constant fields, we first
elucidate which of the generic gauge fields of the form of
eq.~(\ref{Audef}) are consistent with the vacuum Maxwell equations of
eq.~(\ref{Fmunu}). We will take all spatial dependence to reside in
the Kerr--Schild scalar field $\phi\equiv\phi(x,x^i)$, and take the
Kerr--Schild vector to be
\begin{equation}
  k=\sqrt{2}du,
  \label{kudef2}
\end{equation}
such that $k_0=1$ in accordance with the rules of the Kerr--Schild
double copy. We also note that a constant $k^\mu$ is necessarily
geodesic. The only non-vanishing components of the field strength
tensor are then found to be
\begin{equation}
  F_{uv}=-\partial_v\phi=\frac{1}{\sqrt{2}}\partial_x\phi,\quad
  F_{ui}=-\partial_i\phi.
  \label{Fcomps}
\end{equation}
The Maxwell equations impose constraints on the function $\phi$, where
the first of these arise from
\begin{equation}
  \partial_v F^{vu}=0\quad\Rightarrow\quad \partial_x^2\phi=0.
  \label{Maxwell1}
\end{equation}
This in turn implies that we can write
\begin{equation}
  \phi(x,x^i)=\varphi(x^i)+\tilde{\varphi}(x^i)x,
  \label{phixxi}
\end{equation}
for some functions $\varphi(x^i)$ and $\tilde{\varphi}(x^i)$. Next, we
have
\begin{equation}
  \partial_v F^{vi}=0\quad\Rightarrow\quad \partial_x\partial_i\phi=0,
  \label{Maxwell2}
\end{equation}
which has the effect of modifying eq.~(\ref{phixxi}) to the more
specific requirement\footnote{We note for completeness that one could
also include a term linear in the $\{x^i\}$ when solving the
constraint of eq.~(\ref{Maxwell2}). However, one may choose instead to
include this in the function $\varphi(x^i)$, and we will see this
explicitly in what follows.}
\begin{equation}
  \phi(x,x^i)=\varphi(x^i)+fx,\quad f={\rm const.}
  \label{phixxi2}
\end{equation}
Finally, we may consider
\begin{equation}
  \partial_i F^{iv}+\partial_u F^{uv}=0\quad
  \Rightarrow\quad \partial_i^2\varphi=0.
  \label{phixxi3}
\end{equation}
Thus, the gauge field of eq.~(\ref{Audef}), with the Kerr--Schild
vector of eq.~(\ref{kudef2}), is a solution of the vacuum Maxwell
equations provided that the accompanying scalar field has the form of
eq.~(\ref{phixxi2}), with $\varphi(x^i)$ a harmonic function in the
coordinates $\{x^i\}$, which are transverse to the lightcone
coordinate directions. Note also that, when $f\neq 0$, any gauge transformation of the form of eq.~(\ref{thetasol}) can, from eq.~(\ref{phixxi2}), be absorbed in a constant shift of the $x$ coordinate.

Arguably the simplest non-trivial special case of the above
construction is a uniform electric field in the $x$-direction, for
which the harmonic function $\varphi(x^i)$ vanishes. One then has
\begin{equation}
  A=\sqrt{2}fx du,
  \label{Amuconst}
\end{equation}
which leads to a non-zero field strength component
\begin{equation}
  F_{01}=f,
  \label{F01}
\end{equation}
with all others vanishing. This is indeed a constant electric field
$f$ in the positive $x$-direction, as claimed above. Another case that
we will consider in what follows is that of a constant field in the
transverse directions, which corresponds to
\begin{equation}
  \varphi(x^i)=f_i x^i,\quad f=0.
  \label{varphiconst}
\end{equation}
One then finds
\begin{equation}
  A=\sqrt{2} f_i x^i du,\quad F_{0i}=F_{1i}=f_i,
  \label{Amuconst2}
\end{equation}
such that there is a combination of an electric and magnetic
field. The two special cases considered here are complementary, in
that they either involve electric fields aligned with the special
direction $x$, or not. A superposition of these solutions is also
possible, given the linearity of the Maxwell equations. We now
consider the double copy of the above solutions.

\subsection{Double copy of the constant electric field}
\label{sec:DCE}

\begin{figure}
\centering
  \subfloat[][]
             {\scalebox{0.4}{\includegraphics{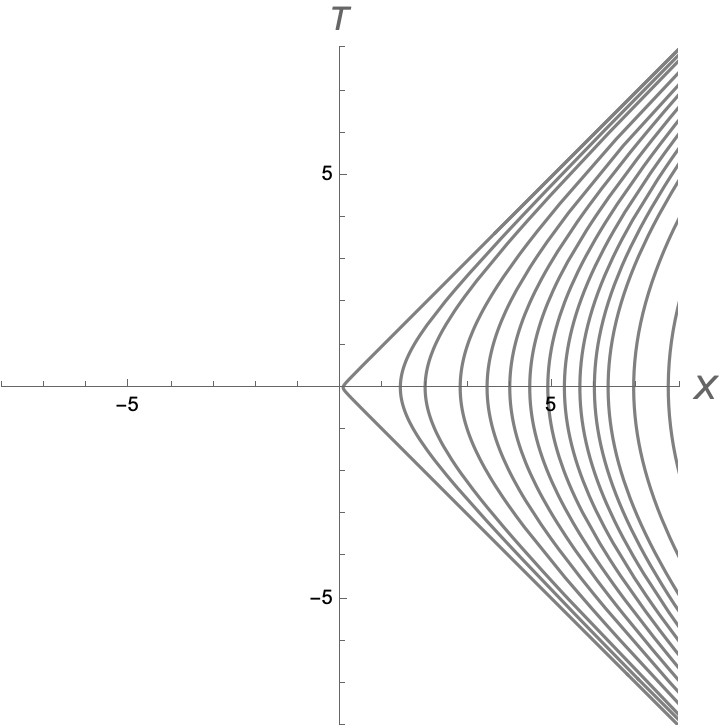}}}
             \qquad\qquad\qquad\qquad
    \subfloat[][]
             {\scalebox{0.4}{\includegraphics{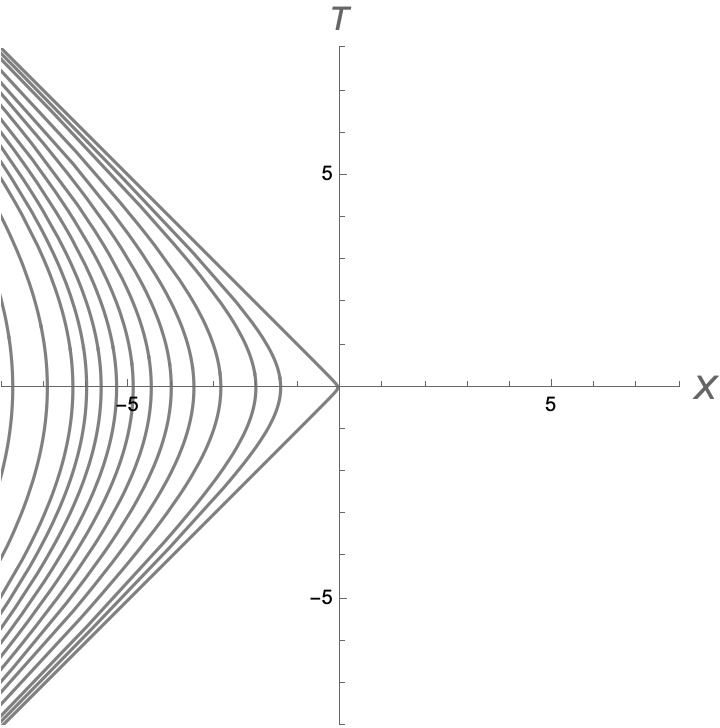}}}
    \caption{Trajectories of objects at rest in the double copy metric
      of eq.~(\ref{ds2E}), as seen in the Minkowski coordinates of
      eqs.~(\ref{dsE2}, \ref{UVdef2}). Results are shown for: (a)
      $f>0$ and fixed values of $h_0<0$; (b) $f<0$ and fixed values of
      $h_0>0$.}
    \label{fig:traj}
\end{figure}

For any static Kerr--Schild gauge field of the form of
eqs.~(\ref{Audef}), the double copy metric is given by
\begin{align}
  g_{\mu\nu}=\eta_{\mu\nu}+\kappa\phi k_\mu k_\nu.
  \label{gmunuKS}
\end{align}
For the specific solution of eq.~(\ref{Amuconst}), this yields a line
element
\begin{align}
  ds^2&=-2dudv+dx_i dx^i+\sqrt{2}f(v-u)du^2\\
  &=-dt^2+dx^2+dx_i dx^i-fx(dx+dt)^2,
  \label{ds2E}
\end{align}
where we have converted to Cartesian coordinates in the second line,
and set the gravitational coupling to one for convenience. As a first
step in interpreting this metric, we can take the Newtonian limit and
consider the gravitational potential
\begin{equation}
  \Phi=-\frac {1}{2}h_{00}=\frac{fx}{2}.
  \label{Phidef}
\end{equation}
We then see that the effect of the graviton is to introduce a uniform
gravitational field, whose strength is precisely related to the
strength of the uniform electric field in the gauge theory. So far so
good, but the apparent non-zero gravitational field implied by
eq.~(\ref{ds2E}) turns out, in fact, to be illusory. To see this, we
may first rescale $u$ and introduce a new coordinate $h$ via\footnote{Note that the freedom to shift $x$ corresponds to the related freedom to shift the $x$ coordinate in the gauge theory, as noted earlier.}
\begin{equation}
  u=t+x,\quad h=-x-\frac{1}{f},
  \label{uhdef}
\end{equation}
so that the metric becomes
\begin{equation}
  ds^2=fhdu^2-2dudh+dx_i dx^i.
\label{ds2Eh}
\end{equation}
Next, one may introduce the coordinates
\begin{align}
  U(u)&=\frac{2}{f}e^{\frac{fu}{2}},\quad 
  V(u,h)=he^{-\frac{fu}{2}},\quad
  X^i(\{x_i\})=x^i,
  \label{UVdef}
\end{align}
in terms of which the metric of eq.~(\ref{ds2E}) becomes
\begin{equation}
  ds^2=-2dUdV+dX_i dX^i.
  \label{dsE2}
\end{equation}
The mapping \eqref{UVdef} is non-invertible, with $U V = 2 f h$ having a fixed sign. One then arrives at the Minkowski metric with Cartesian coordinates
$(T,X,Y,Z)$, and lightcone coordinates
\begin{equation}
  U=\frac{T+X}{\sqrt{2}},\quad V=\frac{T-X}{\sqrt{2}}.
  \label{UVdef2}
\end{equation}
Hence, the apparent gravitational field arising from the double copy
has been completely removed by a coordinate transformation. This tells
us that the original double copy metric of eq.~(\ref{ds2E}) must
correspond to a non-inertial observer, whose apparent gravitational
field is in fact due to a fictitious force. We can verify this
interpretation as follows. First, note that the acceleration of
an observer whose trajectory is described by a curve
$\gamma^\mu(\tau)$ parametrised by proper time $\tau$, is defined by
\begin{equation}
  a^\mu=\frac{d^2\gamma^\mu(\tau)}{d\tau^2}.
    \label{amudef}
\end{equation}
An observer at rest in the non-inertial frame has constant values of
$x$, $h$ and $x^i$, such that one has
\begin{equation}
  u(\lambda)=\lambda,\quad h=h_0={\rm const.},\quad x^i=x_0^i={\rm const.}
  \label{uparam}
\end{equation}
The magnitude of the acceleration is a Lorentz invariant (the {\it
  proper acceleration}) representing the acceleration felt in the
instantaneous rest frame of the observer. We are thus free to
calculate this in any frame, and it is particularly useful to use the
Minkowski frame $(U,V,X^i)$ to simplify the derivatives. The
trajectory in this frame is
\begin{equation}
  \gamma(\lambda)=\left(
  U(\lambda),V(\lambda),X^i(\lambda)
  \right)=\left(
  \frac{2e^{\frac{f\lambda}{2}}}{f},h_0e^{-\frac{f\lambda}{2}},
    x_0^i\right),
    \label{traj1}
\end{equation}
from which one finds
\begin{equation}
  d\tau^2=2dUdV-dX_i dX^i=-fh_0d\lambda^2,
  \label{traj2}
\end{equation}
and thus
\begin{equation}
  d\tau=\pm\sqrt{-fh_0}d\lambda.
  \label{traj3}
\end{equation}
At this point, we take the positive solution so that proper time
increases with increasing parameter $\lambda$. We furthermore assume
$f>0$ so that only observers with $h_0<0$ are physical, as fact that
we will interpret below. The acceleration 4-vector is then
\begin{equation}
  \ddot{\gamma}=-\frac{1}{fh_0}\gamma''(\lambda)
  =\left(\frac{f e^{\frac{f\lambda}{2}}}{2}\partial_U+
  \frac{f^2 h_0}{4}e^{-\frac{f\lambda}{2}}\partial_V\right),
  \label{gammares}
\end{equation}
such that the proper acceleration is found to be
\begin{equation}
  \ddot{\gamma}^2(\tau)=-\frac{f}{4h_0}.
  \label{accel}
\end{equation}
We thus see that in gravity, what was originally the electric field
$f$ now shows up as the acceleration of an observer at rest in the
non-inertial frame picked out by the double copy. To further clarify
the above results, it is instructive to plot the trajectories in
Minkowski coordinates corresponding to fixed values of $h_0<0$. These
are shown in fig.~\ref{fig:traj}(a), and indeed correspond to a
well-known property of uniformly accelerated observers, namely that
their motion in an inertial frame is confined to a {\it Rindler
  wedge}. In the present case, this is a right Rindler wedge due to
the fact that we have taken $f>0$, corresponding to a uniform
acceleration in the positive $X$ direction. Also, the fact that the
wedge emanates from the origin arises from the shifting of the $x$
coordinate in eq.(\ref{uhdef}). The upper and lower boundary of the
wedge in fig.~\ref{fig:traj} constitutes a {\it Rindler
  horizon}\footnote{Throughout this paper, we will adopt the term {\it
  Rindler horizon} for any horizon arising from the effects of
acceleration, as opposed to the more specific meaning of the
particular horizon one sees in standard Rindler coordinates.}, whose
consequence is that a uniformly accelerating particle is causally
disconnected from the upper and lower parts of Minkowski space. One
can, however, cover the left-hand region, by taking $f<0$ and $h_0>0$
above. The resulting trajectories are shown in fig.~\ref{fig:traj}(b),
and fill in a left Rindler wedge. Choices of $f$ and $h_0$ which
violate the requirement $fh_0<0$ would correspond, in principle, to
filling in the upper and lower wedges in Minkowski space. However,
from above we know that this corresponds to imaginary proper time,
which is therefore forbidden.

\begin{figure}
\centering
  \subfloat[][]
             {\scalebox{0.5}{\includegraphics{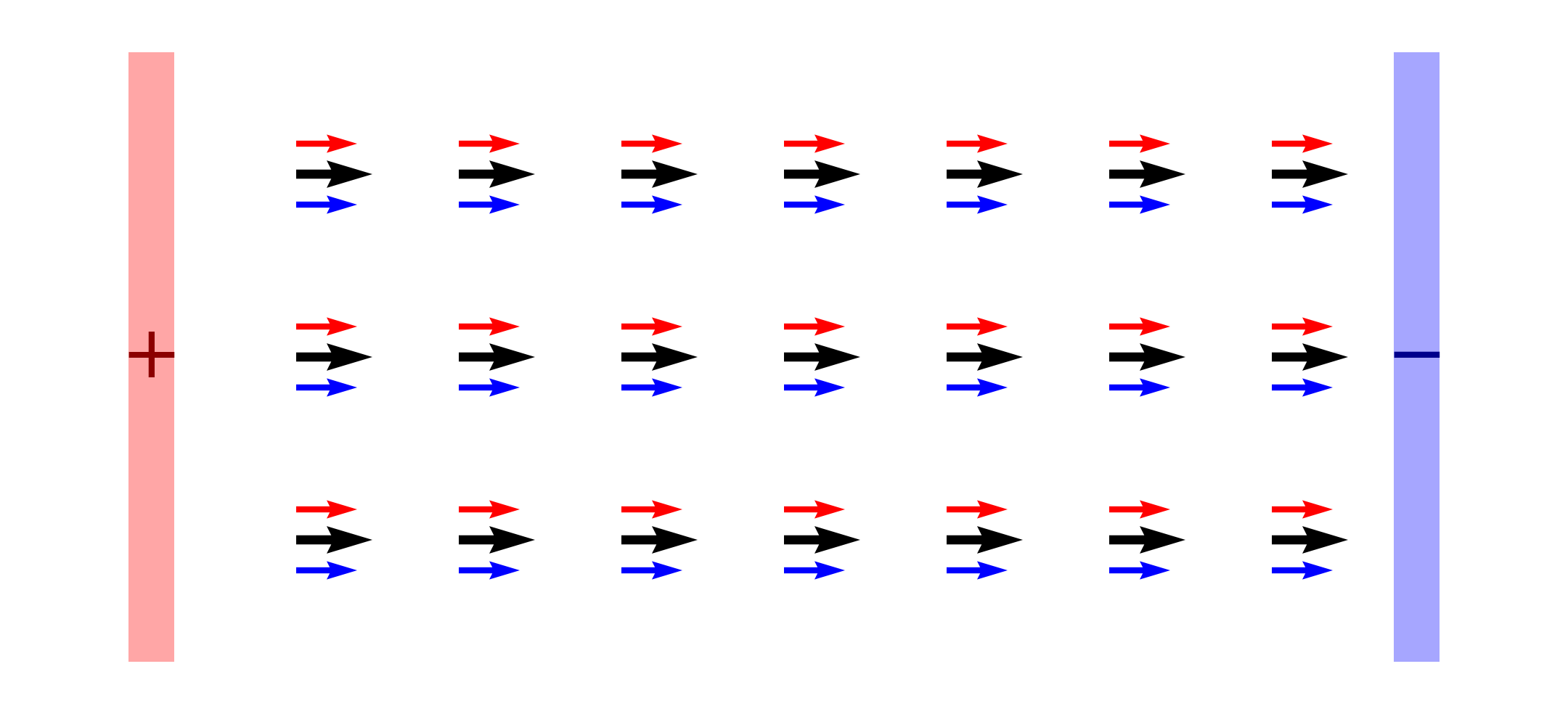}}}
             \qquad\qquad\qquad\qquad
    \subfloat[][]
             {\scalebox{0.5}{\includegraphics{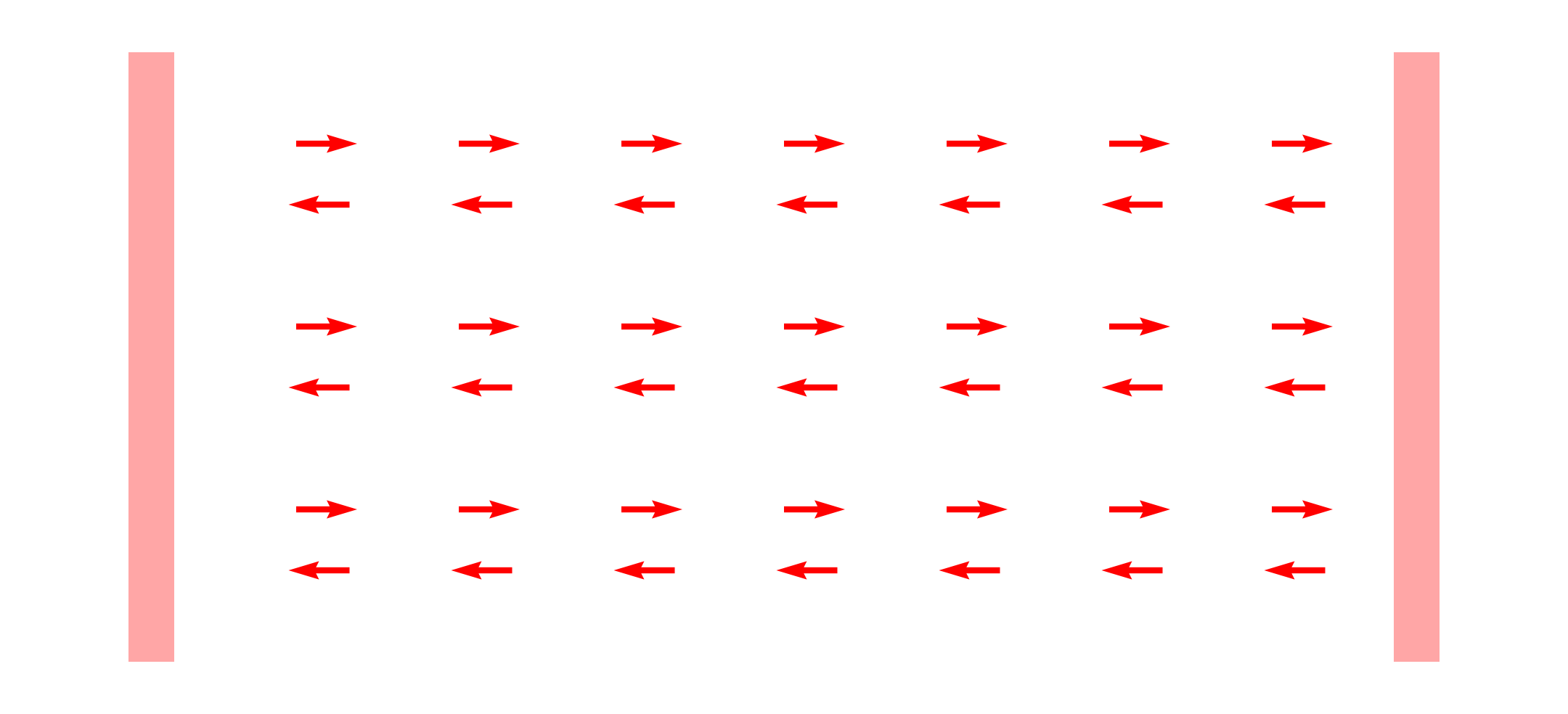}}}
    \caption{(a) A constant electric field can be obtained by placing positively and negatively charged plates at the boundary of space, where the fields due to each plate (shown in red and blue) combine to make the black arrows; (b) the double copy maps the charged plates to masses, whose gravitational fields (shown in red) cancel out.}
    \label{fig:capacitor}
\end{figure}

There is a potentially interesting interpretation of the above double
copy, which helps explain why the double copy of a constant electric
field does not give rise to a genuine gravitational field. In
electrostatics, one can ``source'' a uniform electric field filling
all space by placing two oppositely charged plates at the boundary of
space, as shown in fig.~\ref{fig:capacitor}(a), and such that the
entire spatial volume can be thought of as forming the interior of a
giant capacitor. The uniform electric field can then be thought of as
arising from the superposition of the electric fields due to each
plate. The double copy replaces charge with mass, leading to the
situation shown in fig.~\ref{fig:capacitor}(b), in which two equally
massive plates are at the spatial boundary. Unlike in electrodynamics,
however, gravitational ``charge'' is universally attractive, and thus
the gravitational fields from each massive plate cancel each other
out. What makes this interesting from a double copy point of view is
that there is no local charge information giving rise to the fields on
the gauge and gravity sides of the correspondence. Rather, the
information needed to encode the fields is in the boundary conditions
of the spacetime. Nevertheless, the double copy somehow knows to map
these boundary conditions appropriately, and this adds to our
intuition of how it works in practice.

Whilst the above picture has a certain charm, it is perhaps not to be
taken too literally. In the coordinate system of eq.~(\ref{ds2E})
picked out by the double copy, there is indeed a remnant of the
electric field from the gauge theory, in that it appears as the
acceleration of the non-inertial frame. Furthermore, the gravity
solution knows whether the gauge theory has boundary conditions
corresponding to fig.~\ref{fig:capacitor}(a), or the opposite
situation in which the charged plates are swapped. The latter case
corresponds to reversing the sign of the electric field, and hence of
$f$. This then gives rise, in gravity, to the left Rindler wedge of
fig.~\ref{fig:traj}(b), rather than the right Rindler wedge of
fig.~\ref{fig:traj}(a).

\subsection{Double copy of the constant electromagnetic field}
\label{sec:DCEB}

\begin{figure}
  \begin{center}
    \scalebox{0.5}{\includegraphics{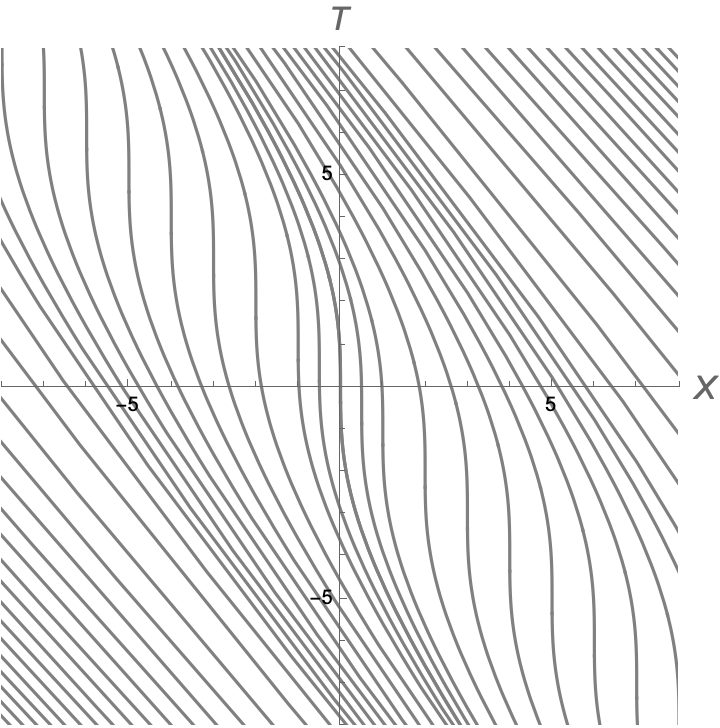}}
    \caption{Trajectories in the $(X,T)$ plane associated with
      stationary observers in the non-inertial frame of
      eq.~(\ref{ds2EB}). Results are shown for different initial
      values of $x=x_0$, with $f^2=1$ and $x^i_0=0$.}
    \label{fig:trajEB1}
  \end{center}
\end{figure}

\begin{figure}
\centering
  \subfloat[][]
             {\scalebox{0.4}{\includegraphics{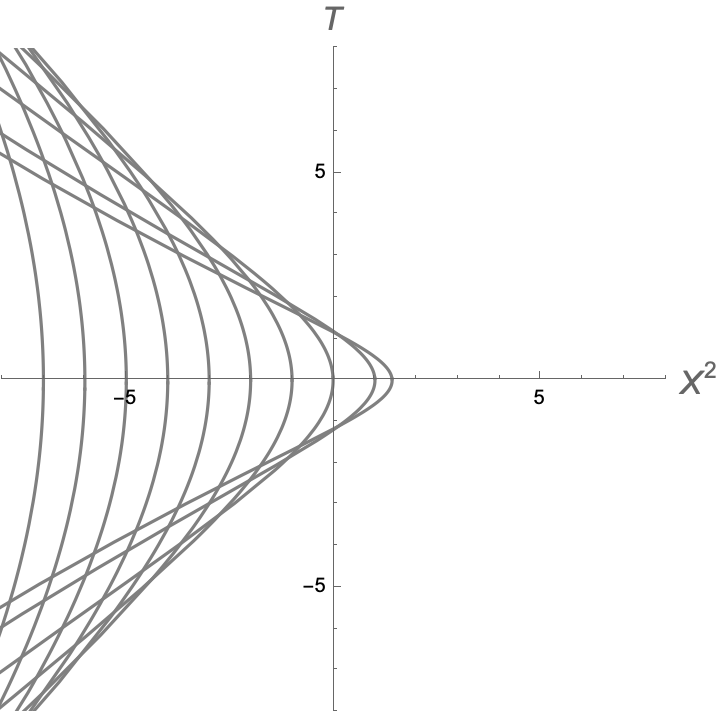}}}
             \qquad\qquad\qquad\qquad
    \subfloat[][]
             {\scalebox{0.4}{\includegraphics{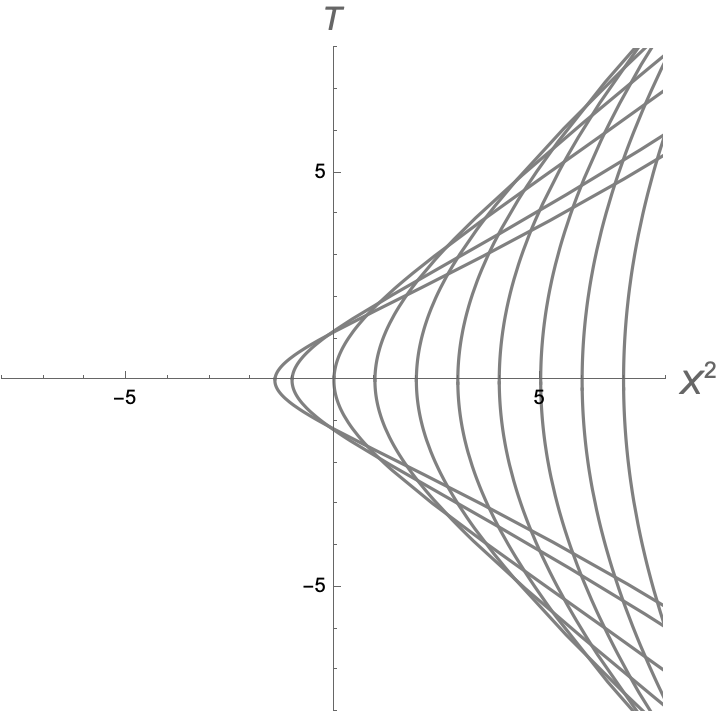}}}
    \caption{Trajectories in the $(X^2,T)$ plane associated stationary
      observers in the non-inertial frame of
      eq.~(\ref{ds2EB}). Results are shown for different initial
      values $x^2=x^2_0$, and $x_0=0$, with: (a) $f^2=1$; (b)
      $f^2=-1$.}
    \label{fig:trajEB2}
\end{figure}

Our second special case of gauge fields whose Kerr--Schild vector is of
the form of eq.~(\ref{kudef2}) was that of eq.~(\ref{Amuconst2}). The
gauge field is linear in the transverse coordinates $x^i$, giving rise
to both a constant electric and magnetic field. The double copy is
straightforwardly obtained, and leads to the line element
\begin{equation}
  ds^2=-2dudv+dx_i dx^i+\sqrt{2} f_i x^i du^2.
  \label{ds2EB}
\end{equation}
The Newtonian limit in this case gives rise to a gravitational
potential
\begin{equation}
  \Phi=-\frac{f_i x^i}{\sqrt{2}},
  \label{PhiEB}
\end{equation}
and hence an apparent constant gravitational field in the transverse
directions, whose direction is dictated by the vector $\vec{f}$
corresponding to the electric field in the gauge theory. Once again,
though, one can in fact remove this gravitational field through a
coordinate transformation, such that the double copy metric amounts to
a non-inertial frame. In the present case, the metric of
eqs.~(\ref{dsE2}, \ref{UVdef2}) is obtained upon defining
\begin{align}
  U(u)&=u,\quad V(u,v,\{x^j\})=v-\frac{u f_i x^i}{\sqrt{2}}+
  \frac{|\vec{f}|^2 u^3}{12},\quad X^i(\{x_j\})=x^i-
  \frac{f^i u^2}{2\sqrt{2}}.
  \label{UVEB}
\end{align}
Differently from the previous case this map is one-to-one, i.e. into and onto Minkowski space.
We can obtain the proper acceleration by once again considering the
trajectory, in the Minkowski frame, of an observer at rest in the
non-inertial frame, where the latter will be characterised by
\begin{equation}
  u(\lambda)=\lambda,\quad x=x_0={\rm const.}, \quad
  v=\lambda-2\sqrt{x_0},\quad x^i=x^i_0={\rm const.}
  \label{observer2}
\end{equation}
The trajectory in the Minkowski frame is then
\begin{equation}
  \gamma^\mu(\lambda)=\left(
  U(\lambda),V(\lambda),X^i(\lambda)
  \right)=
  \left(
  \lambda,\lambda-\sqrt{2} x_0-\frac{\lambda f_i x^i_0}
         {\sqrt{2}}+\frac{|\vec{f}|^2\lambda^3}{12},
         x_0^i-\frac{f^i\lambda^2}{2\sqrt{2}}
         \right),
         \label{trajEB}
\end{equation}
from which one finds a proper time
\begin{equation}
  d\tau = \sqrt{2-\sqrt{2} f_i x^i_0}d\lambda.
  \label{dtauEB}
\end{equation}
We have again taken the positive branch of the square root, so that
the parameter $\lambda$ increases with increasing proper time of the
observer. The presence of the square root then tells us that only
certain trajectories are possible in Minkowski space, namely those for
which $f_ix^i_0<\sqrt{2}$. As in the previous section, this is due to
a Rindler horizon associated with the acceleration, which we will
examine in more detail below. From eqs.~(\ref{trajEB}, \ref{dtauEB}),
the acceleration of the observer in the Minkowski frame is
\begin{equation}
  \ddot{\gamma}(\lambda(\tau))=\frac{1}{2-\sqrt{2}f_i x_0^i}
  \left(
  \frac{|\vec{f}|^2\lambda(\tau)}{2}\partial_V-\frac{f_i}{\sqrt{2}}
  \partial_{X^i}\right),
  \label{accelEB}
\end{equation}
from which the proper acceleration is
\begin{equation}
  \ddot{\gamma}^2=\frac{|\vec{f}|^2}{4(\sqrt{2}-f_i x^i_0)}.
  \label{propaccelEB}
\end{equation}
As in the previous section, we see that the electric field in the
gauge theory dictates the acceleration on the gravity side of the
double copy. Here, the electric field is in the transverse directions
in the gauge theory, which is reflected in the fact that the proper
acceleration depends only on the transverse vector $\vec{f}$. This is
also reflected in the Rindler horizon above, which involves only the
transverse coordinates $\{x^i\}$. To visualise the latter, let us take
the vector $\vec{f}$ to lie in the $x^2$
direction. Figure~\ref{fig:trajEB1} shows the Minkowski trajectories
corresponding to stationary observers in the double copy frame, for
$x^2=0$, and different values of $x_0$. One sees that there is no
horizon in this plane, as expected given that the acceleration is in
the transverse directions. Figure~\ref{fig:trajEB2}(a) shows
trajectories projected into the $(X^2,T)$ plane, for $f^2=1$ and
$x_0=0$, and we can clearly see the Rindler horizon associated with
the fact that there is now a constant acceleration in the $X^2$
direction. The different shape of this horizon arises from the more
complicated shape of the trajectories given that there is a magnetic
field in addition to the electric field in the gauge theory. Similar
results for $f^2=-1$ are shown in fig.~\ref{fig:trajEB2}(b).

\section{String motion in background fields}
\label{sec:strings}

In the previous section, we have introduced particular gauge fields
corresponding to constant electric or magnetic fields, and interpreted
their double copies. We now turn to how to use these results to
examine the double copy within the context of classical string
theory. As stated in the introduction, our motivation stems from the
well-known KLT relations~\cite{Kawai:1985xq} describing closed string
scattering amplitudes as sums of products of open string scattering
amplitudes, and whose low energy limit is the tree-level double copy
relating gluons (from open strings), and gravitons (from closed
strings, which also give rise to dilaton and axion amplitudes). For
the open string amplitudes, we may consider the open strings as test
objects on a vacuum (zero) electromagnetic background. Similarly, the
closed strings are test objects on a vacuum (Minkowski) spacetime
background. Given that the field theory double copy has been extended
to non-trivial gauge and gravity backgrounds (see
e.g. refs.~\cite{Bahjat-Abbas:2017htu,Carrillo-Gonzalez:2017iyj,Lipstein:2019mpu,Lipstein:2023pih,Ilderton:2025gug,Ilderton:2024oly,Adamo:2020qru}),
it also then makes sense to study the motion of closed strings on a
non-trivial gravity background, as a double copy of open strings
moving in the gauge background obtained as a single copy of the
appropriate graviton. We will here restrict ourselves to classical
motion of a single open or closed string, which amounts -- in
scattering amplitude terms -- to considering a 2-point function in the
presence of a background field. We will ignore backreaction effects of
the probe strings on the gauge or gravity backgrounds themselves. Our
goal is to ascertain how the strings on each side of the double copy
correspondence behave on their respective backgrounds, and to use this
behaviour to gain intuition about this particular extension of the
classical double copy.

Let us first address how our results from section~\ref{sec:gauge} can
be embedded in a string theory context. Electromagnetic fields are
supported by D-branes, where non-trivial fields are sourced by open
strings that carry charges at their endpoints. In the absence of open
strings, D-branes can contain constant electromagnetic fields,
consistent with the fact that these are a vacuum solution of the
relevant field equations. For the latter, we must replace the Maxwell
Lagrangian of electromagnetism with the {\it DBI Lagrangian}
\begin{equation}
  {\cal L}_{\rm DBI}=\sqrt{-{\rm det}({\bf \eta}+2\pi\alpha' {\bf F})}.
  \label{LDBI}
\end{equation}
Here matrix notation is used in the determinant, where ${\bf \eta}$
and ${\bf F}$ are the Minkowski metric and field strength
respectively. The action arising from eq.~(\ref{LDBI}) is the low
energy effective action for open strings interacting with the D-brane,
where $\alpha'=1/(2\pi T_s)$ is related to the string tension
$T_s$. In other words, it describes the dynamics of massless open
string modes (photons) in the limit of slowly varying fields. The
field equations obtained from eq.~(\ref{LDBI}) are:
\begin{equation}
  \partial^\nu {F_\mu}^\lambda\left(1-(2\pi\alpha' {\bf F})^2
  \right)^{-1}_{\lambda\nu}=0,
  \label{DBIEOM}
\end{equation}
from which it immediately follows the constant background
electromagnetic fields we examined in section~\ref{sec:gauge} are
consistent solutions of the full DBI theory, hence can act as
background gauge fields for strings, where one assumed the presence of
a space-filling D-brane. An analogous story holds on the gravity side
of the double copy, where the Einstein-Hilbert action for General
Relativity will be supplemented by higher-order terms in the string
tension $\alpha'$, involving local invariants constructed from the
Riemann curvature. However, for the constant gauge fields considered
in section~\ref{sec:gauge}, the double copy metrics were all such that
the Riemann curvature vanished (i.e. they amounted to Minkowski space
written in non-inertial coordinates). Hence, the double copy metrics
obtained above are also exact string backgrounds, and we can 
consider the motion of open and closed strings in the gauge and
gravity backgrounds respectively.

The behaviour of an open string in a constant electromagnetic field is
well-known~\cite{Fradkin:1985qd,Burgess:1986dw,Nesterenko:1989pz}, and
its equations of motion arise from the action
\begin{equation}
  S=-\frac{1}{4\pi\alpha'}\int_{\Sigma}d^2\sigma \left[
    \dot{X}^2-{X'}^2
    \right]+\int_{\partial \Sigma}d\tau A_\mu(X)\dot{X}^\mu.
  \label{Sstring}
\end{equation}
Here the first term on the right-hand side constitutes the Polyakov
action for the open string in conformal gauge, where the integration
is over the string worldsheet $\Sigma$ with coordinates $\tau$ and
$\sigma$. The second term describes the coupling of the open string
endpoints to the gauge field, such that the integration is over the
worldsheet boundary $\partial\Sigma$. In what follows, we choose units
such that
\begin{equation}
  \alpha'=\frac{1}{2\pi}.
  \label{alpha'}
\end{equation}
Then the equations of motion are
\begin{align}
  &\Box X^\mu=0
  \label{stringEOM1}
\end{align}
and
\begin{equation}
  \partial_\sigma X^\mu+{F^\mu}_\nu \partial_\tau X^\nu=0,\quad
  \sigma=0,\pi.
  \label{stringEOM2}
\end{equation}
For the case of the constant electric field considered in
section~\ref{sec:DCE}, the field strength is
\begin{equation}
  F_{\mu\nu}=\left(
  \begin{array}{ccccc}
    0 & f & 0 & \ldots &  0\\
    -f & 0 & 0 & \ldots & 0\\
    0 & 0 & 0 & \ldots & 0\\
    \vdots & \vdots & \vdots & \vdots & \vdots\\
        0 & 0 & 0 & \ldots & 0
  \end{array}
  \right),
  \label{FmunuE}
\end{equation}
so that the equations of motion simplify to
\begin{align}
  T'&=-f\dot{X},\quad \sigma=0,\pi;\notag\\
  X'&=-f\dot{T},\quad \sigma=0,\pi;\notag\\
  \Box X^\mu&=0,
  \label{stringEOM3}
\end{align}
where the dot and prime correspond to differentiation with respect to
$\tau$ and $\sigma$ respectively. One may decouple the equations by
switching to lightcone coordinates according to eq.~(\ref{UVdef2}):
\begin{align}
  U'&=-f\dot{U},\quad \sigma=0,\pi;\notag\\
  V'&=f\dot{V},\quad \sigma=0,\pi;\notag\\
  \Box X^i&=\Box U=\Box V=0.
  \label{stringEOM4}
\end{align}
The harmonic condition for $U(\tau,\sigma)$ implies the general
solution
\begin{equation}
  U(\tau,\sigma)=L(\tau+\sigma)+R(\tau-\sigma),
  \label{Usol}
\end{equation}
such that the first equation in eq.~(\ref{stringEOM4}) yields
\begin{equation}
  R'(\xi)=\frac{1+f}{1-f}L'(\xi).
  \label{instab1}
\end{equation}
We can clearly see the presence of an instability as $f\rightarrow 1$
which, in our choice of units, amounts to the condition $f< T_s$ in
terms of the string tension. The physical interpretation of this
instability is that the electric field $f$ is such as to pull the
(charged) endpoints of the open string away from each other. Once the
electric field becomes strong enough, it can overcome the string
tension, such the open string becomes unstable and decays. Another way
to see this is to note that the determinant appearing in the DBI
action of eq.~(\ref{LDBI}) is given (in our choice of units) by
\begin{align}
  {\rm det}({\bf \eta}+{\bf F})=
{\rm det}\left[\left(
  \begin{array}{ccccc}
    -1 & f & 0 & \ldots &  0\\
    -f & 1 & 0 & \ldots & 0\\
    0 & 0 & 1 & \ldots & 0\\
    \vdots & \vdots & \vdots & \vdots & \vdots\\
        0 & 0 & 0 & \ldots & 1
  \end{array}
  \right)\right] = f^2-1.
  \label{det}
\end{align}
Positivity of the argument of the square root in eq.~(\ref{LDBI}) then
imposes the condition $f^2<1$, such that the electric field must be
less than a critical value. Above this, the DBI action becomes
imaginary, which indicates an unstable situation, and indeed precisely
the instability described above.

So much for the constant electric field in the $X$ direction. We may
also examine the general case of a constant field strength, as
considered in section~\ref{sec:DCEB}:
\begin{equation}
  F_{\mu\nu}=\left(
  \begin{array}{ccccc}
    0 & f & f_2 & \ldots &  f_{D-1}\\
    -f & 0 & f_2 & \ldots & f_{D-1}\\
    -f_2 & -f_2 & 0 & \ldots & 0\\
    \vdots & \vdots & \vdots & \vdots & \vdots\\
        -f_{D-1} & -f_{D-1} & 0 & \ldots & 0
  \end{array}
  \right).
  \label{FmunuEB}
\end{equation}
An explicit calculation then gives
\begin{equation}
  {\rm det}({\bf \eta}+{\bf F})=f^2-1,
  \label{det2}
\end{equation}
leading to the same condition $f^2<1$ as before. To see what is going
on, let us introduce a book-keeping parameter $\alpha$, and write the
argument of the determinant as
\begin{align}
  {\rm det}({\bf \eta}+{\bf F})&=
  {\rm det}\left[\left(
  \begin{array}{ccccc}
    -1 &  f & \alpha f_2 & \ldots &  \alpha f_{D-1}\\
    -f & 1 & f_2 & \ldots & f_{D-1}\\
    -\alpha f_2 & -f_2 & 1 & \ldots & 0\\
    \vdots & \vdots & \vdots & \vdots & \vdots\\
        -\alpha f_{D-1} & -f_{D-1} & 0 & \ldots & 1
  \end{array}
  \right)\right]\notag\\
  &=f^2-1+\sum_{i=2}^{D-1}(\alpha^2-1)f_i^2.
  \label{det3}
\end{align}
We thus see that, for the transverse directions, the electric and
magnetic contributions to the DBI actions occur with opposite signs,
and precisely cancel when the physical value $\alpha=1$ is
reinstated. This leaves the contribution of the electric field along
the $X$ direction, which has no magnetic counterpart. Thus, the only
instability is associated with an electric field in the X direction.

Returning to the purely electric case, the mode expansions for the
first two string coordinates (with our choice of units for $\alpha'$)
are
\begin{align}
  T(\tau,\sigma)&=t_0+\frac{1}{\pi}\frac{1}{1-f^2}(p^0\tau+f p^1\sigma)
  +i\sqrt{\frac{1}{\pi}}\sum_{n\neq 0}\frac{1}{n}\alpha_n^0 e^{-in\tau}
  \cos(n\sigma);\notag\\
  X(\tau,\sigma)&=x_0+\frac{1}{\pi}\frac{1}{1-f^2}(p^1\tau+fp^0\sigma)
  +i\sqrt{\frac{1}{\pi}}
  \sum_{n\neq 0}\frac{1}{n}\alpha_n^1e^{-in\tau}\cos(n\sigma),
\label{modeexp1}
\end{align}
as may be verified by direct substitution in eq.~(\ref{stringEOM3}). Equation~(\ref{modeexp1}) may be compared with the mode expansions in the absence of a background field:
\begin{align}
  T(\tau,\sigma)&=t_0+\frac{p^0 \tau}{\pi}
  +i\sqrt{\frac{1}{\pi}}\sum_{n\neq 0}\frac{1}{n}\alpha_n^0 e^{-in\tau}
  \cos(n\sigma);\notag\\
  X(\tau,\sigma)&=x_0+\frac{p^1\tau}{\pi}
  +i\sqrt{\frac{1}{\pi}}
  \sum_{n\neq 0}\frac{1}{n}\alpha_n^1e^{-in\tau}\cos(n\sigma).
\label{modeexp2}
\end{align}
We then see that the non-zero background gauge field does not affect the
oscillator modes, but leads to a modification of the zero modes, with
the instability as $f\rightarrow 1$ clearly visible.

Having examined the behaviour of the open string in a gauge theory
background, we can now examine the double copy situation of a closed
string moving in the background metrics explored in
section~\ref{sec:gauge}. A suitable starting point is the action for
the closed string in a background metric $g_{\mu\nu}$:
\begin{equation}
  S=\frac{1}{4\pi\alpha'}\int d\tau d\sigma \sqrt{-\gamma}
  \,\gamma^{ab}\,g_{\mu\nu}\,\partial_a X^\mu\, \partial_b X^\nu.
  \label{Scurved}
\end{equation}
Here $\gamma_{ab}$ is the worldsheet metric with determinant $\gamma$,
and the latin indices $(a,b)$ denote worldsheet
coordinates. Evaluating the action for the double copy metric of
eq.~(\ref{ds2Eh}), one finds
\begin{align}
  S&=\frac{1}{4\pi\alpha'}\int d\tau d\sigma \sqrt{-\gamma}
  \Big[\gamma^{\tau\tau}(-2\dot{U}\dot{H}+fH\dot{U}^2+\dot{X}_i\dot{X}^i)
    +2\gamma^{\tau\sigma}(-\dot{U}H'-U'\dot{H}+fH\dot{U}U'+{X'}_i{X'}^i)
    \notag\\
    &\quad
    +\gamma^{\sigma\sigma}(-2U'H'+fH(U')^2+{X'}_i{X'}^i)\Big].
\label{SstringE}
\end{align}
Note that, following convention, we have labelled spacetime coordinate
fields on the string worldsheet by capital letters, corresponding to
the lower case letters appearing in eq.~(\ref{ds2Eh}). These are not
to be confused with the capital letter coordinates used for the
non-inertial (Minkowski) frames in section~\ref{sec:gauge}. To fix the
gauge freedom in choosing the worldsheet metric, we may adopt
lightcone gauge
\begin{equation}
  U(\tau)=\tau,
  \label{LCgauge1}
\end{equation}
with the supplementary conditions (see
e.g. ref.~\cite{Polchinski:1998rq})
\begin{equation}
  \gamma=-1,\quad \partial_\sigma\gamma_{\sigma\sigma}=0,\quad
  \gamma_{\tau\sigma}(\tau,0)=0.
  \label{gammaconds}
\end{equation}
This particular gauge fixing has the advantage that cubic and higher
terms in the string action do not appear. Solving also for the inverse
worldsheet metric $\gamma^{ab}$ in terms of $\gamma_{ab}$,
eq.~(\ref{SstringE}) becomes
\begin{equation}
  S=\frac{1}{4\pi\alpha'}\int d\tau d\sigma \Big(\gamma_{\sigma\sigma}
  (-fh+2\dot{h}-\dot{X}_i\dot{X}^i)-2\gamma_{\tau\sigma}
  (\tilde{H}'-\dot{X}_i{X'}^i)+\gamma_{\sigma\sigma}^{-1}
  (1-\gamma_{\tau\sigma}^2)X'_i{X'}^i\Big),
  \label{SstringE2}
\end{equation}
where we have defined
\begin{equation}
  H(\tau,\sigma)=h(\tau)+\tilde{H}(\tau,\sigma),\quad
    h(\tau)=\frac{f}{l}\int_0^l d\sigma H(\tau,\sigma).
    \label{Htildedef}
\end{equation}
That is, we have decomposed the coordinate field $H(\tau,\sigma)$ into
its mean value $h(\tau)$ integrated along the string at a given
worldsheet time $\tau$ (where $l$ is the string length), and a
deviation $\tilde{H}(\sigma,\tau)$ which has zero mean value. The
action may then be simplified to
\begin{align}
  S&=\frac{l}{2\pi\alpha'}
\int d\tau \gamma_{\sigma\sigma}\left(\dot{h}-\frac{fh}{2}\right)
\notag\\
&\quad+\frac{1}{4\pi\alpha'}
\int d\tau d\sigma \Big(-\gamma_{\sigma\sigma}\dot{X}_i\dot{X}^i
-2\gamma_{\tau\sigma}
  (\tilde{H}'-\dot{X}_i{X'}^i)+\gamma_{\sigma\sigma}^{-1}
  (1-\gamma_{\tau\sigma}^2)X'_i{X'}^i\Big),
  \label{SstringE3}
\end{align}
and we may then note that $\gamma_{\tau\sigma}$ and $\tilde{H}$ appear
as auxiliary (non-dynamical) fields. Their equations of motion
respectively imply
\begin{equation}
  \tilde{H}'=\dot{X}_i{X'}^i-\gamma_{\tau\sigma}
  \gamma_{\sigma\sigma}^{-1}{X'}_i{X'}^i,\quad
  \partial_\sigma \gamma_{\tau\sigma}=0.
  \label{EOMS1}
\end{equation}
The second equation, together with the boundary condition
$\gamma_{\tau\sigma}(\tau,0)=0$ and the fact that $\sigma$ is defined
on a periodic interval, implies
\begin{equation}
  \gamma_{\tau\sigma}=0,
  \label{gam0}
\end{equation}
and substituting this plus the first condition in eq.~(\ref{EOMS1})
into eq.~(\ref{SstringE3}) yields
\begin{equation}
  S=\frac{l}{2\pi\alpha'}\int d\tau
  \gamma_{\sigma\sigma}\left(\dot{h}-\frac{fh}{2}\right)
  +\frac{1}{4\pi\alpha'}\int d\tau d\sigma \Big(
  -\gamma_{\sigma\sigma}\dot{X}_i\dot{X}^i+\gamma^{-1}_{\sigma\sigma}
  X'_i{X'}^i
  \Big).
  \label{SstringE4}
\end{equation}
The equation of motion for $h$ implies
\begin{equation}
  \dot{\gamma}_{\sigma\sigma}=-\frac{f}{2}\gamma_{\sigma\sigma}\quad
  \Rightarrow\quad
  \gamma_{\sigma\sigma}(\tau)=\bar{\gamma}e^{-\frac{f\tau}{2}},
  \label{gammasol}
\end{equation}
where $\bar{\gamma}$ is a constant. To make further progress, it is
convenient to decompose the transverse component fields in terms of
(non-)zero mode contributions, analogous to eq.~(\ref{Htildedef}):
\begin{equation}
  X^i(\tau,\sigma)=x^i(\tau)+\tilde{X}^i(\tau,\sigma),
  \label{Xitildedef}
\end{equation}
and such that eq.~(\ref{SstringE4}) becomes
\begin{align}
  S=\frac{l}{2\pi\alpha'}\int d\tau \gamma_{\sigma\sigma}
  \left(\frac{fh}{2}-\dot{h}\right)+
  \frac{1}{4\pi\alpha'}\int d\tau d\sigma\Big(
  \gamma_{\sigma\sigma}\left(
  \dot{x}_i\dot{x}^i+\dot{\tilde{X}}_i\dot{\tilde{X}}^i
  \right)
  -\gamma^{-1}_{\sigma\sigma}\tilde{X}'_i({\tilde{X}'})^i
  \Big).
  \label{SstringE5}
\end{align}
We are now in a position to obtain the mode expansions for the various
coordinate fields. The equation of motion for $\gamma_{\sigma\sigma}$
gives a particularly cumbersome equation for the zero mode $h$, which
we do not focus on explicitly here. The form of the transverse zero
modes in our choice of gauge is obtained from the equation of motion
for $x^i$:
\begin{equation}
  \partial_\tau(\gamma_{\sigma\sigma}\dot{x}^i)=0\quad\Rightarrow\quad
  \dot{x}^i=\dot{x}^i(0)e^{\frac{f\tau}{2}},
  \label{xisol}
\end{equation}
where we have used eq.~(\ref{gammasol}). Finally, the equation of
motion for the transverse non-zero modes gives
\begin{equation}
  \ddot{\tilde{X}}_i-\frac{f}{2}\dot{\tilde{X}}_i=\frac{e^{f\tau}}
       {\bar{\gamma}^2}\tilde{X}''.
       \label{tildeXeq}
\end{equation}
Substituting the general periodic form
\begin{equation}
  \tilde{X}^i(\tau,\sigma)=\sum_{n\neq 0}x^i_n(\tau)e^{\frac{2\pi in\sigma}
    {l}}
  \label{Xiexpand}
\end{equation}
gives an equation for each Fourier modes $x_n^i$:
\begin{equation}
  \ddot{x}_n^i-\frac{f}{2}\dot{x}^i_n=
  \left(\frac{2\pi i n}{\bar{\gamma} l}\right)^2e^{f\tau}x_n^i,
  \label{xnisol}
\end{equation}
whose solution is
\begin{equation}
  x_n^i(\tau)=\alpha_n^i\exp\left(\frac{4\pi i n}{\bar{\gamma}fl}e^{f\tau}
    \right)
    +\tilde{\alpha}_n^i\exp\left(-\frac{4\pi i n}{\bar{\gamma}fl}e^{f\tau}
    \right).
      \label{xnisol2}
\end{equation}
We thus see that, in lightcone gauge with gauge fixing as above, the
closed string moving in the double copy metric of eq.~(\ref{ds2Eh})
``knows'' about the background's origin as a double copy of a constant
electric field, through its mode expansion. Alternatively, one could
transform coordinates to an inertial frame, in which case one would
be back to the standard situation of a closed string moving in
Minkowski spacetime. Then, however, the boundary conditions of the
spacetime would be different, as discussed in section~\ref{sec:gauge},
due to the presence of a Rindler horizon. Whichever way one looks at
things, there is always a remnant of the double copy having taken
place, and a sense in which the closed string knows about the gauge
theory calculation. Notably, a counterpart of the instability as
$f\rightarrow 1$ (in units of the string tension) does not appear for
the closed string. This well-known fact is due to the absence of
endpoints for a closed string, which can carry charge. One thus finds
a qualitatively different behaviour in the closed string situation
compared to its single copy open string analogue, and this is not the
first time that such effects have arisen in the double copy
literature. In general, it is known that the double copy for field
theory amplitudes or classical solutions can make pathological
behaviour vanish, or indeed appear. Canonical examples include the
removal of certain infrared divergences when double-copying gauge
theory to gravity~\cite{Oxburgh:2012zr}, and the worsening of UV
behaviour~\cite{Bern:2008qj,Bern:2010ue,Bern:2010yg}. 

For completeness, let us also note that one can can carry out a
similar analysis for the case of the double-copied constant electric
and magnetic field described in section~\ref{sec:DCEB}, where the
analogue of eq.~(\ref{SstringE}) is then
\begin{align}
  S&=\frac{1}{4\pi\alpha'}=\int d\tau d\sigma\sqrt{-\gamma}
  \Big[\gamma^{\tau\tau}\left(
    -2\dot{U}\dot{V}+\sqrt{2}f_iX^i\dot{U}^2+\dot{X}_i\dot{X}^i
    \right)\notag\\
    &\quad+2\gamma^{\tau\sigma}\left(
    -\dot{U}V'-U'\dot{V}+\sqrt{2}f_iX^i\dot{U}U'+X'_i\dot{X}^i
    \right)\notag\\
    &\quad+\gamma^{\sigma\sigma}\left(
    -2U'V'+\sqrt{2}f_iX^i(U')^2+X'_i{X'}^i
    \right)\Big].
  \label{SstringEB}
\end{align}
Imposing the same lightcone gauge fixing conditions as
eq.~(\ref{LCgauge1}, \ref{gammaconds}), we may also decompose each
coordinate to isolate the zero mode:
\begin{equation}
  V(\tau,\sigma)=v(\tau)+\tilde{V}(\tau,\sigma),\quad
  X^i(\tau,\sigma)=x^i(\tau)+\tilde{X}^i(\sigma,\tau).
  \label{Vdecomp}
\end{equation}
The equation of motion for $\tilde{V}$ implies $\gamma_{\tau\sigma}=0$
as before. Furthermore, the equation for $v$ turns out to imply
$\gamma_{\sigma\sigma}={\rm const.}$, in contrast to the exponential
behaviour found in the previous example. The action then reduces to
\begin{equation}
  S=\frac{1}{4\pi\alpha'}\int d\tau d\sigma
  \Big[
    -\gamma_{\sigma\sigma}\left(
    -2\dot{v}+\sqrt{2}f_ix^i+\dot{x}_i\dot{x}^i+\dot{\tilde{X}}_i
    \dot{\tilde{X}}^i
    \right)+\gamma^{-1}_{\sigma\sigma}\tilde{H}'_i{\tilde{H}}'^i
    \Big].
  \label{SstringEB2}
\end{equation}
The equation of motion for $x^i$ is
\begin{equation}
  \ddot{x}_i=\frac{f_i}{\sqrt{2}},
  \label{xisolEB}
\end{equation}
leading to a quadratic behaviour for $x_i$. For the transverse
oscillator modes, one ultimately finds
\begin{equation}
  \tilde{X}^i(\tau,\sigma)=\sum_{n\neq 0}x^i_n(\tau) e^{\frac{2\pi in\sigma}
    {l}},\quad
  x_n^i(\tau)=\alpha_n^i\exp\left(
  \frac{2\pi in\tau}{\gamma_{\sigma\sigma}l}
  \right)
  +\tilde{\alpha}_n^i\exp\left(-
  \frac{2\pi in\tau}{\gamma_{\sigma\sigma}l}
  \right).
  \label{tildeXisolEB}
\end{equation}
The behaviour seen here is in stark contrast to the exponential
behaviour that is seen in the zero modes and mode expansion for the
purely electric case. This is not surprising, however, given that a
similar contrast is already seen in the Minkowski trajectories of
particles at rest in the non-inertial frame picked out from the double
copy (eqs.~(\ref{traj1}, \ref{trajEB})). The fact that, for the
electric case, exponential behaviour manifests also in the behaviour
of the transverse zero modes is a consequence of the gauge fixing.

\section{Discussion}
\label{sec:discuss}

In recent years, the double copy has become a widely studied research
area linking scattering amplitudes and classical solutions in gauge
and gravity theories. For tree-level amplitudes, the double copy has a
well-known string theoretic origin in the KLT relations between open
and closed string scattering. Given the possibility of double-copying
classical backgrounds, one may then expand the remit of the double
copy by examining how classical strings behave in appropriate gauge
and gravity backgrounds.

In this paper, we have performed a case study involving arguably the
simplest classical backgrounds, namely those involving constant
electric and magnetic fields in the gauge theory. We constructed the
double copy of these backgrounds using the well-known Kerr--Schild
approach, finding that the double copy picks out a non-inertial
reference frame in all cases, such that the apparent gravitational
field vanishes upon transforming to Minkowski coordinates. However, in
the latter case the Minkowski spacetime still ``knows'' about its
origin as a double copy of a gauge background, in that observers at
rest in the double copy frame will produce a Rindler horizon in the
Minkowski frame. We offered a physical picture for this, involving the
replacement of charge-related boundary conditions in the gauge theory,
with mass-related boundary conditions in the gravity theory.

Motivated by the KLT relations for amplitudes, we examined closed
strings in a gravitational background as the double copy of open
strings interacting with a background gauge field, where the latter is
obtained as the classical single copy of the closed string's
background spacetime. The advantage of using constant field strength
tensors is that the resulting fields in both gauge theory and gravity
can be shown to be exact solutions in string theory, including all
$\alpha'$ corrections. For the particular constant electric and
magnetic fields considered, we find that closed strings in the double
copy frame know about the background's origin from gauge theory,
through the appearance of the electric field in the mode expansion for
strings. If one were to transform instead to a Minkowski frame, the
mode expansion would be conventional, but the double copy would still
show up through the boundary conditions on the spacetime itself.

Understanding new situations related by the double copy, including how
the physics in the gauge and gravity theories manifests itself,
provides new insights behind the origin and scope of this fascinating
correspondence. We hope that our results inspire others to consider
the interplay of the double copy with classical or quantum string
theory on non-trivial backgrounds, and look forward to more work in
this area.

\section*{Acknowledgments}

CDW and MRA are supported by the UK Science and Technology Facilities
Council (STFC), including through the Consolidated Grant ST/P000754/1
``String theory, gauge theory and duality''.


\bibliography{refs}
\end{document}